\documentclass[12pt]{article}
\usepackage{graphicx}
\usepackage{array,dcolumn}      
\usepackage{axodraw}
\begin{document}
%
%
%
%
\begin{titlepage}
\topmargin 0in
\begin{center}
\hfill  { DATE} \marginpar{CHANGE}
\end{center}
\vskip 0.25in
\begin{center}{\Large\bf
 THE ONE-LOOP QED IN NONCOMMUTATIVE SPACE  }
 \vskip .2in
{\bf Nguyen tien Binh}
 \vskip .1in
Universit\'e de Neuch\^atel \\
CH--2000 Neuch\^atel, Switzerland
\end{center}
\vskip .2in
\begin{center}
{\bf Abstract}
\end{center}
\begin{quote}
Returning to the old problems in ordinary QED,by an appropriate
extension of the dimensional regularization method in
noncommutative space we try to provide a quite coherent look into
NCQED.The renormalisation of theories, the $\beta$ function,the
vacuum polarisation of photon,the general structure of vertex
fermion-photon ,the anomalous magnetic moment (AMM) of fermions
and the validity of Ward identity at the one-loop level are
reinvestigated.
\end{quote}
\end{titlepage}
\setcounter{footnote}{0}
\newpage
%
%
\section{INTRODUCTION}
Since 1999 the space-time non-commutativity has been realized from
string theory when open string propagates in the presence of
constant background antisymmetric tensor field [1].In recent years
the noncommutative field theories have generated a lot of
interests in many aspects (see reference in [14] ) ,for instance,
the Lorenz invariance ,the unitarity
[12,16,17,20],instanton[13],Standard model on NCM space-time and
then the new NCM theories[15] based on the extension of the basic
relation which indicates that space-time loses its condition of
continuum:
\begin{equation}
[\widehat{x}_{\mu},\widehat{x}_{\nu}]=i\theta_{\mu\nu}
\end{equation}
In spite of that, there are some old aspects, particularly in
NCMQED,which have to be reinvestigated.In the works [5],[8],[18]
the structure of the vertex is considered,the Ward identity have
called attention in [5], particularly  [21] is an explicit
investigation for its validity in the two processes
$e^{+}e^{-}\rightarrow \gamma\gamma$ and $\gamma\gamma\rightarrow
\gamma\gamma$,Now, in this work we try to investigate the same
problems in the NCMQED at one-loop level by using the dimensional
regularization and perturbative method. For the pedagogical
purpose , we will organize the paper in the following way. :Sec.2
is devoted to remind some principal properties of NCMQED,Sec.3 is
reserved to present some results for the renormalisation of
theories,Sec.4 is a brief consideration of the structure of vertex
,Sec.5 for checking the validity of Ward identity ,Sec.6 is the
discussion of the AMM and we will finally end up with some
comments and remarks.
\section{PERTURBATIVE THEORY OF NCMQED}
1.\hspace{0.2in}Let us begin with the pure U(1) NCM Yang-Mills
action in space- time dimension d:
\begin{equation}
S_{YM}=-\frac{1}{4}\int d^{d}x F_{\mu\nu}(x)\star F^{\mu\nu}(x)
\end{equation}
Here the $\star$-product is defined as:
\begin{equation}
f(x)\star
g(x)=e^{\frac{i}{2}\theta^{\mu\nu}\partial_{\mu}^{(\xi)}\partial_{\nu}^{(\eta)}}
f(x+\xi)g(x+\eta)\biggr|_{\xi,\eta\rightarrow 0}\\
\end{equation}
$\theta^{\mu\nu}$ is antisymmetric matrix which has the dimension
of area. To avoid problems with unitarity we will assume that only
the space-space components of $\theta^{\mu\nu}$ are non
zero,namely, $ \theta^{0\nu}=0 $.The field strength
$F_{\mu\nu}(x)$ is defined by:
\begin{equation}
F_{\mu\nu}(x)=\partial_{\mu}A_{\nu}(x)-\partial_{\nu}A_{\mu}(x)-ig[A_{\mu}(x),A_{\nu}(x)]_{M}
\end{equation}
and the Moyal bracket is:
\begin{equation}
[f,g]_{M}=f\star g-g\star f
\end{equation}
Note that: even in U(1) case $A_{\mu}$ couples to itself since the
field strength $F_{\mu\nu}$ has the non-linear term in $A_{\mu}$.
Assuming that the fields decrease so promptly at infinity that the
space -time integral of a Moyal bracket vanishes.It is easily to
show that the action (2) is invariant under the gauge
transformations following: \begin{equation} A_{\mu}(x)\rightarrow
A'_{\mu}(x)=U(x)\star A_{\mu}(x)\star
U^{-1}(x)+\frac{i}{g}U(x)\star
\partial_{\mu}U^{-1}(x)
\end{equation}
 Where:
$$U(x)=\left(e^{i\lambda(x)}\right)_{\star}=1+i\lambda(x)+\frac{(i)^{2}}{2}\lambda(x)\star\lambda(x)+...$$
$$ U^{-1}(x)=\left(e^{-i\lambda(x)}\right)_{\star}=1-i\lambda(x)+\frac{(-i)^{2}}{2}\lambda(x)\star\lambda(x)+...$$
$$ U(x)\star U^{-1}(x)=\textbf{1}$$
Now, we consider the fundamental representation of the matter in
 which the covariant derivative is defined as:
 $$ D_{\mu}\psi(x)=\partial_{\mu}\psi(x)+ig\psi(x)\star A_{\mu}(x)$$
In terms of ordinary product the action for U(1) Yang-Mills fields
and the matter field can be rewritten as :
\begin{equation}
S_{YM-Matter}=\int
d^{4}x\left\{-\frac{1}{4}F^{\mu\nu}F_{\mu\nu}+\overline{\psi}i\partial\hspace{-0.2cm}/\psi
-ge^{\frac{i}{2}p\wedge
p'}\overline{\psi}A\hspace{-0.2cm}/\psi-m\overline{\psi}\psi\right\}
\end{equation}
In order to obtain the non-singular free
 propagator for the gauge fields we need to introduce the gauge-fixing term.We shall do this in a consistent
 way by using the BRS formalism.Let us introduce the
 ghost fields $ c,\overline{c}$ and the auxiliary fields B and
 define the BRST transformations as follow:
$$ \delta_{B}A_{\mu}(x)=D_{\mu}c(x)=\partial_{\mu}c(x)-i[A_{\mu}(x),c(x)]_{M}$$
$$\delta_{B}c(x)=-\left(\overline{c}(x)\star c(x)\right)$$
$$\delta_{B}\overline{c}(x)=B(x)$$
 \begin{equation}
  \delta_{B}B(x)=0
 \end{equation}
 is nilpotent.\\
 To keep track of the renormalisation of the composite
 transformations $\delta_{B}A_{\mu}(x) $ and $\delta_{B}c(x) $
 one also introduces the external fields $J_{\mu}(x)$ and $ H(x)$
 which couple to them.
 The Faddev-Popov and gauge fixing action is given as:
 \begin{equation}
 S_{GF-ghost}=\int
 d^{d}x\left\{-\frac{1}{2\xi}\partial_{\mu}A^{\mu}(x)\star\partial_{\nu}A^{\nu}(x)
 +\partial^{\mu}\overline{c}(x)\star\left(\partial_{\mu}c(x)-i[A_{\mu}(x),c(x)]_{M}\right)\right\}
 \end{equation}
 Where $\xi$ is the gauge fixing parameter.\\
 and the action for the sources is defined by:
 \begin{equation}
S_{sources}=\int d^{d}x\left\{K^{\mu}\star A_{\mu}+J\star
B+(\overline{\eta}\star c-\overline{c}\star\eta)+H\star(c\star
c)+J^{\mu}\star D_{\mu}c\right\}
\end{equation}
 It is easily to show that$ S_{GF-ghost}$ is invariant under the BRST
 transformations. The complete action for NCQED in a general covariant gauge is
 given as:
 \begin{equation}
S_{tot}=S_{YM}+S_{Matter}+S_{gf-ghost}+S_{sources}=S_{inv}+S_{sources}
 \end{equation}
 where:
 $S_{YM+Matter}$ and $S_{gf-ghost}$ are defined in (7),(10)
  respectively.
As a result, the propagator is the same as in the commutative
  counterpart but each vertex will accompanies with a phase factor
  which depends on the momenta outgoing from the vertex and the consequence
  is that the Feynman rules in Feynman gauge are derived and presented in figure
  (1).
\newpage
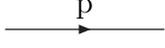
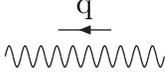
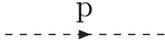
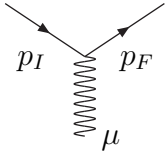
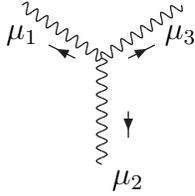
\begin{center}
\begin{picture}(600,600)(0,0)
\ArrowLine(10,590)(70,590)\Text(40,600)[]{p}
\Text(140,590)[]{\hspace{5cm}
$iS_{(p)}=\frac{i}{p\hspace{-0.12cm}/-m+i\epsilon}$}
\Photon(10,550)(70,550){4}{10}
\ArrowLine(50,560)(30,560)\Text(40,570)[]{q}
\Text(140,550)[]{\hspace{5cm}$iD^{\mu\nu}=\frac{-ig^{\mu\nu}}{q^{2}+i\epsilon}$}
\DashArrowLine(10,510)(70,510){3}\Text(40,520)[]{p}
\Text(140,510)[]{\hspace{5cm}$iD(p)=\frac{i}{p^{2}+i\epsilon}$}
\ArrowLine(10,470)(40,450)\Photon(40,450)(40,420){4}{8}
\ArrowLine(40,450)(70,470) \Text(20,450)[]{$p_{I}$}
\Text(60,450)[]{$p_{F}$}\Text(50,420)[]{$\mu$}
\Text(140,450)[]{\hspace{5cm}$=ie\gamma^{\mu}e^{\frac{i}{2}p_{I}\wedge
p_{F}}$}
\Photon(10,380)(40,360){2}{8} \Photon(40,360)(70,380){2}{8}
\Photon(40,360)(40,320){2}{8}
\ArrowLine(50,360)(60,365)\ArrowLine(30,360)(20,365)\ArrowLine(50,340)(50,330)
 \Text(10,370)[]{$\mu_{1}$}\Text(70,370)[]{$\mu_{3}$}\Text(50,315)[]{$\mu_{2}$}
 \Text(120,340)[]{\hspace{5cm}$=2esin(\frac{1}{2}p_{1}\wedge
 p_{2})$}
 \Text(120,320)[]{\hspace{7cm}$[(p_{1}-p_{2})^{\mu3}g^{\mu1\mu2}+(p_{2}-p_{3})^{\mu1}g^{\mu2\mu3}
 +(p_{3}-p_{1})^{\mu2}g^{\mu3\mu1}]$}
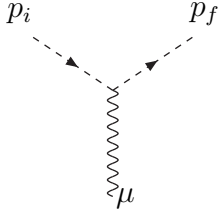
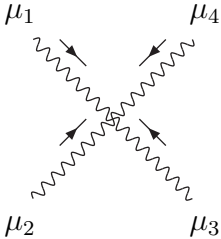
\DashArrowLine(10,280)(40,260){3}\DashArrowLine(40,260)(70,280){3}\Photon(40,220)(40,260){2}{8}
\Text(120,260)[]{\hspace{7cm}$=2iep^{\mu}_{f}sin(\frac{p_{i}\wedge
 p_{f}}{2})$}
\Text(5,290)[]{$p_{i}$}\Text(45,220)[]{$\mu$}\Text(75,290)[]{$p_{f}$}
\Photon(10,180)(40,150){2}{8}\Photon(10,120)(40,150){2}{8}
\Photon(40,150)(70,180){2}{8}\Photon(70,120)(40,150){2}{8}
\ArrowLine(20,180)(30,170)\ArrowLine(20,140)(30,150)\ArrowLine(60,140)(50,150)\ArrowLine(60,180)(50,170)
 \Text(5,110)[]{$\mu_{2}$}\Text(5,190)[]{$\mu_{1}$}\Text(75,110)[]{$\mu_{3}$}\Text(75,190)[]{$\mu_{4}$}
\Text(120,150)[]{\hspace{7cm}$=4ie^{2}\biggl[\sin(\frac{p_{1}\wedge
 p_{2}}{2})\sin(\frac{p_{3}\wedge
 p_{4}}{2})$}
 \Text(120,130)[]{\hspace{7cm}$(g^{\mu1\mu3}g^{\mu2\mu4}-g^{\mu1\mu4}g^{\mu2\mu3})$}
 \Text(120,110)[]{\hspace{7cm}$+sin(\frac{p_{3}\wedge p_{1}}{2})sin(\frac{p_{2}\wedge
 p_{4}}{2})$}
 \Text(120,90)[]{\hspace{7cm}$(g^{\mu1\mu4}g^{\mu2\mu3}-g^{\mu1\mu2}g^{\mu3\mu4})$}
 \Text(120,70)[]{\hspace{7cm}$+sin(\frac{p_{1}\wedge p_{4}}{2})sin(\frac{p_{2}\wedge
 p_{3}}{2})$}
 \Text(120,50)[]{\hspace{7cm}$(g^{\mu1\mu2}g^{\mu3\mu4}-g^{\mu1\mu3}g^{\mu2\mu4})\biggr]$}
 \Text(100,30)[]{Fig (1): The Feynman rules for NCMQED}
  \end{picture}
\end{center}
\begin{center}
\begin{picture}(550,550)(0,0)
\ArrowLine(140,220)(220,220)
\PhotonArc(180,220)(30,0,180){3}{20}\ArrowLine(110,220)(140,220)\ArrowLine(220,220)(250,220)
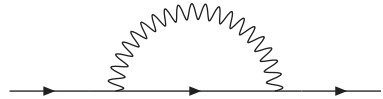
\Text(150,180)[]{(Figure 2):The correction to electron
self-energy}
\DashCArc(90,510)(30,0,360){3}\Photon(20,510)(60,510){3}{4}\Photon(120,510)(160,510){3}{4}
\Text(90,460)[]{(figure 3.a)}
\PhotonArc(290,510)(30,0,360){3}{40}\Photon(220,510)(260,510){3}{4}\Photon(320,510)(360,510){3}{4}
\Text(290,460)[]{(figure 3.b)}
\PhotonArc(90,410)(30,0,360){3}{40}\Photon(20,410)(60,410){3}{8}\Photon(120,410)(160,410){3}{8}
\Text(90,360)[]{(figure 3.c)}
 \ArrowArc(290,410)(30,360,180) \ArrowArc(290,410)(30,180,0)\Photon(220,410)(260,410){3}{6}\Photon(320,410)(360,410){3}{6}
\Text(290,360)[]{(figure 3.d)}
 \Text(150,320)[]{Fig (3): The correction to vacuum polarization of photon}
\ArrowLine(20,50)(50,70)\ArrowLine(50,70)(80,90)\Photon(80,130)(80,90){4}{8}
\ArrowLine(80,90)(110,70)\ArrowLine(110,70)(140,50)\Photon(50,70)(110,70){4}{8}
\ArrowLine(70,60)(90,60)\ArrowLine(90,120)(90,100)\Text(25,60)[]{p}\Text(135,70)[]{p'}\Text(50,85)[]{p-k}
\Text(110,85)[]{p'-k} \Text(110,110)[]{q=p'-p}\Text(80,50)[]{k}
\Text(80,30)[]{diagram(a)}\Text(250,30)[]{diagram(b)}
\ArrowLine(190,50)(220,70)\ArrowLine(220,70)(280,70)\ArrowLine(280,70)(310,50)
\ArrowLine(260,120)(260,100)\Photon(220,70)(250,90){4}{6}
\Photon(250,90)(280,70){4}{6}\Photon(250,130)(250,90){4}{6}
\Text(280,110)[]{q=p'-p}\Text(215,85)[]{k-p}\Text(290,85)[]{p'-k}\Text(250,60)[]{k}\Text(195,60)[]{p}\Text(315,60)[]{p'}
\Text(150,0)[]{Fig (4): The correction to $
\psi\overline{\psi}A_{\mu}$-vertex}
\end{picture}
\end{center}
  \newpage
\section{THE RENORMALISED NCMQED}
It is useful to note that firstly we work in the Feynman gauge and
secondly,in the following calculations we assume that the Bessel
functions are finite.
\part*{{\large\bf 1. The one-loop fermion self-energy:}}
  The electron self-energy receives one-loop correction through
  only one diagram (2):
\begin{equation}
\widetilde{\sum}_{(p)}=-ie^{2}\mu^{\epsilon}\int\frac{d^{d}k}{(2\pi)^{d}}
\gamma_{\nu}\frac{1}{(p\hspace{-0.2cm}/-k\hspace{-0.2cm}/)}\gamma^{\nu}\frac{1}{k^{2}}
\end{equation}
  Noting in this diagram that the
  phase factors accompanied with the two vertices cancel with each
  other.Thus the contribution is the same as in ordinary QED .So,
  not only UV divergence can be subtracted by the usual rescaling
  of wave function and electron's mass but also the finite part
  does not change .In the dimensional renormalisation [22], the counter term for the fermion loop is defined as:
  $$ Z_{2}=1+\Delta_{2}$$
  where $\Delta_{2}$ is decomposed into two parts: infinite and finite part which in the on-shell condition are given by:
  $$\Delta_{2}^{\infty}=-\frac{e^{2}}{16\pi^{2}}\left(\frac{2}{\epsilon}-\gamma_{E}+\right)$$
  and,
  \begin{equation}
  \Delta_{2}^{F}=-\frac{e^{2}}{(4\pi)^{2}}\left(2-\ln\frac{m^{2}}{\mu^{2}}-4\int_{0}^{1}dz\frac{1-z^{2}}{z}\right)
  \end{equation}
  So, the renormalisation constant is :
  \begin{equation}
  Z_{2}=1-\frac{e^{2}}{8\pi^{2}\epsilon}
  \end{equation}
  \part*{{\large\bf 2. The one-loop photon self-energy.}}
  The one-loop photon self-energy  receives contribution from four
  diagram (3a)-(3d) in figure (3).\\
  (i)\hspace{0.5cm}The diagram (3d) is the contribution of the electron loop.It
  turns out that its contribution is  the same as in ordinary QED
  since like the electron self-energy the two phase factors at
  two vertices are cancelled with each other. We get:
  \begin{equation}
i\Pi^{(d)}_{\mu\nu}(p)=\frac{ie^{2}N_{f}}{6\pi^{2}\epsilon}(p_{\mu}p_{\nu}-\eta_{\mu\nu}p^{2})
=\frac{ie^{2}}{(4\pi)^{2}\epsilon}\frac{-8N_{f}}{3}(\eta_{\mu\nu}p^{2}-p_{\mu}p_{\nu})
  \end{equation}
  Where $ N_{f}$ denote the number of independent fields with
  charge $\pm 1$. \\
  (ii)\hspace{0.5cm}For three diagrams
  (3a),(3b) and (3c).In the dimensional regularization we can
  rewrite the contribution of these diagrams in the form:
  $$ i\Pi_{\mu\nu}^{(i)}=e^{2}\mu^{4-d}C^{(i)}\int
  \frac{d^{d}k}{(2\pi)^{d}}
  \frac{1-cosp\wedge k}{k^{2}(p+k)^{2}}\textit{N}_{\mu\nu}^{(i)}$$
  where i=a,b,c
$$ C^{(a)}=\frac{1}{2},\hspace{0.2in}C^{(b)}= -1,\hspace{0.2in}C^{(c)}=\frac{1}{2}$$
and:
$$ \textit{N}^{(a)}_{\mu\nu}=4\eta_{\mu\nu}(d-1)(p+k)^{2}$$
$$ \textit{N}^{(b)}_{\mu\nu}=2k_{\mu}(p+k)_{\nu}$$
$$ \textit{N}^{(c)}_{\mu\nu}=2\left[\eta_{\mu\nu}(5p^{2}+2pk+2k^{2})+k_{\mu}k_{\nu}(4d-6)+p_{\mu}p_{\nu}(d-6)
+(p_{\mu}k_{\nu}+p_{\nu}k_{\mu})(2d-3)\right]$$ The contribution
of three diagrams is:
\begin{equation}
i\Pi_{\mu\nu}^{(abc)}=e^{2}\mu^{4-d}\int
  \frac{d^{d}k}{(2\pi)^{d}}
  \frac{1-cosp\wedge k}{k^{2}(p+k)^{2}}\textit{N}_{\mu\nu}^{(abc)}
  \end{equation}
  where
  $$\textit{N}_{\mu\nu}^{(abc)}=\Sigma_{i}C^{(i)}\textit{N}_{\mu\nu}^{(i)}$$
  $$  =2d\eta_{\mu\nu}k^{2}+4(d-2)k_{\mu}k_{\nu}+(2d+3)\eta_{\mu\nu}p^{2}+(d-6)p_{\mu}p_{\nu}$$
  $$+2(2d-1)\eta_{\mu\nu}kp
  +(2d-3)p_{\mu}k_{\nu}+(2d-5)p_{\nu}k_{\mu} $$
  Using the dimensional regularization,we see that the photon self-energy receives the contributions
  from the planar part and the non-planar part:
  \begin{equation}
i\Pi_{\mu\nu}^{(abc)}=i\Pi_{\mu\nu(planar)}^{(abc)}+i\Pi_{\mu\nu(non-planar)}^{(abc)}
  \end{equation}
  where:
\begin{equation}
i\Pi_{\mu\nu(planar)}^{(abc)}=e^{2}\mu^{4-d}\int_{0}^{1}dz\int
  \frac{d^{d}k}{(2\pi)^{d}}
  \frac{\textit{N}_{\mu\nu}^{(abc)}}{[k^{2}-M^{2}]^{2}}
  \end{equation}
    \begin{equation}
i\Pi_{\mu\nu(non-planar)}^{(abc)}=-e^{2}\mu^{4-d}\int_{0}^{1}dz\int
  \frac{d^{d}k}{(2\pi)^{d}}
  (cosp\wedge k)\frac{\textit{N}_{\mu\nu}^{(abc)}}{[k^{2}-M^{2}]^{2}}
  \end{equation}
For the planar part in the limit $ \epsilon\rightarrow 0$we get:
$$i\Pi_{\mu\nu(planar)}^{(abc)}=\frac{ie^{2}}{(4\pi)^{2}}
\biggl\{\frac{20}{3\epsilon}(\eta_{\mu\nu}p^{2}-p_{\mu}p_{\nu})\biggr\}$$
\begin{equation}
-\frac{ie^{2}}{(4\pi)^{2}}\int_{0}^{1}dz\left(\ln\frac{M^{2}}{4\pi\mu^{2}}\right)\left(20\eta_{\mu\nu}M^{2}
+\textit{M}_{\mu\nu}\right)-\frac{ie^{2}}{(4\pi)^{2}}\frac{10}{3}\eta_{\mu\nu}p^{2}
\end{equation}
For the non-planar part,we obtain:
$$i\Pi_{\mu\nu(np)}^{(abc)}=-\frac{ie^{2}}{(4\pi)^{2}}\int_{0}^{1}dz\left(\frac{4\pi\mu^{2}}{-M^{2}}\right)^{\frac{\epsilon}{2}}$$
$$\biggl\{\left(2d^{2}+4(d-2)\right)\eta_{\mu\nu}(-M^{2})\left(\frac{Z}{2}\right)^{\frac{\epsilon}{2}-1}K_{\frac{\epsilon}{2}-1}(z)$$
$$+(-M^{2})\left(d\eta_{\mu\nu}(\widetilde{p})^{2}+2(d-2)\widetilde{p_{\mu}}\widetilde{p_{\nu}}\right)
\left(\frac{Z}{2}\right)^{\frac{\epsilon}{2}-2}K_{\frac{\epsilon}{2}-2}(z)$$
\begin{equation}+2\textit{M}_{\mu\nu}\left(\frac{Z}{2}\right)^{\frac{\epsilon}{2}}K_{\frac{\epsilon}{2}}(z)\biggr\}
  \end{equation}
  where$$ Z=|\widetilde{p}|M $$
Noting that there is no divergence in  Bessel functions even when
$\epsilon\rightarrow 0$ .So, the infinite terms in the
contribution of three diagrams (3a)-(3c) come from the planar
part. With the above result,summing (15),(20) and(21) we get the
one-loop photon self-energy :
$$i\Pi_{\mu\nu}^{(abcd)}(p)=i\Pi_{\mu\nu}^{(d)}(p)+i\Pi_{\mu\nu}^{(abc)}(p)$$
\begin{equation}
i\Pi_{\mu\nu}^{(abcd)}(p)=\frac{ie^{2}}{(4\pi)^{2}\epsilon}\left(\frac{20}{3}
-\frac{8N_{f}}{3}\right)\left(\eta_{\mu\nu}p^{2}-p_{\mu}p_{\nu}\right)+finite
\end{equation}
The renormalisation constant $Z_{3}$ is:
\begin{equation}
Z_{3}=1+\frac{e^{2}}{16\pi^{2}\epsilon}\left(\frac{20}{3}-\frac{8N_{f}}{3}\right)
\end{equation}
\part*{{\large\bf 3.Contribution of diagram (4a)}}
Now, we consider the vertex presented in figure (4). In
d-dimensions the proper vertex is :
\begin{equation}
\Lambda_{\mu}^{(a)}(p,q,p')
=-ie^{2}\mu^{\epsilon}e^{\frac{i}{2}p\wedge p'}\int
\frac{d^{d}k}{(2\pi)^{d}}e^{-ik\tilde{q}}\left[\gamma_{\nu}\frac{1}{p'\hspace{-0.25cm}/-k\hspace{-0.25cm}/-m}\gamma_{\mu}
\frac{1}{p\hspace{-0.2cm}/-k\hspace{-0.25cm}/-m}\gamma^{\nu}\frac{1}{k^{2}-
m^{2}_{\gamma}}\right]
\end{equation}
\ where: $m_{\gamma}$ is the mass of photon,$\epsilon=4-d$ and
$k\tilde{q}=k\wedge q=k^{\mu}\theta^{\mu\nu}q_{\nu}$\\
 By standard Feynman parameterisation , we get,
$$ \Lambda^{(a)}_{\mu}(p,q,p')=\frac{2e^{2}\mu^{\epsilon}}{(4\pi)^{\frac{d}{2}}}e^{\frac{i}{2}p\wedge p'}
\int_{0}^{1}dx\int_{0}^{1-x}dye^{-i(x+y)p\wedge p'}$$
$$\biggl\{\gamma_{\mu}\biggl[\frac{(2-d)^{2}}{2(M^{2}_{a})^{\frac{\epsilon}{2}}}
\left(\frac{Z_{a}}{2}\right)^{\frac{\epsilon}{2}}K_{\frac{\epsilon}{2}}(Z_{a})
-\frac{(2-d)}{4}\frac{(\widetilde{q})^{2}}{(M^{2}_{a})^{\frac{\epsilon}{2}-1}}
\left(\frac{Z_{a}}{2}\right)^{\frac{\epsilon}{2}-1}K_{\frac{\epsilon}{2}-1}(Z_{a})\biggr]$$
$$+\frac{(2-d)}{2}\frac{\widetilde{q}_{\mu}\widetilde{q\hspace{-0.20cm}/}}{(M^{2}_{a})^{\frac{\epsilon}{2}-1}}
\left(\frac{Z_{a}}{2}\right)^{\frac{\epsilon}{2}-1}K_{\frac{\epsilon}{2}-1}(Z_{a})$$
\begin{equation}
-\frac{N_{2a}}{(M^{2}_{a})^{\frac{\epsilon}{2}+1}}
\left(\frac{Z_{a}}{2}\right)^{\frac{\epsilon}{2}+1}K_{\frac{\epsilon}{2}+1}(Z_{a})\biggr\}
\end{equation}
 In the limit $d=4$ :
$$  \Lambda^{(a)}_{\mu}(p,q,p')=\frac{2e^{2}}{(4\pi)^{2}}e^{\frac{i}{2}p\wedge p'}
\int_{0}^{1}dx\int_{0}^{1-x}dye^{-i(x+y)p\wedge p'}$$
$$\biggl\{\gamma_{\mu}\biggl[2K_{0}(Z_{a})
+\frac{(\widetilde{q})^{2}(M^{2}_{a})}{Z_{a}}K_{-1}(Z_{a})\biggr]$$
\begin{equation}
-\left(\widetilde{q}_{\mu}\widetilde{q}\hspace{-0.20cm}/\right)
\left(\frac{2M_{a}^{2}}{Z_{a}}\right)K_{-1}(Z_{a})
-\frac{N_{2a}}{(M^{2}_{a})^{2}}
\left(\frac{Z_{a}}{2}\right)K_{1}(Z_{a})\biggr\}
\end{equation}\\\\
Since we can safely set $\epsilon=0$ (there is no pole)with the
Bessel functions, it is easy to see that the contribution of
diagram (4a) is finite.
\part*{{\large\bf 4.Contribution of diagram (4b)}}
For diagram (4b).Starting with:
$$\Lambda_{\mu}^{(b)}(p,q,p')
=-ie^{2}\mu^{\epsilon}e^{\frac{i}{2}p\wedge p'}\int
\frac{d^{d}k}{(2\pi)^{d}}\left(e^{ik\wedge q}e^{-ip\wedge
p'}-1\right)$$
$$\left[\gamma^{\rho}\frac{k\hspace{-0.20cm}/+m}{(k^{2}-m^{2})}
\gamma^{\nu}\frac{1}{(p'-k)^{2}-m^{2}_{\gamma}}\frac{1}{(p-k)^{2}-m^{2}_{\gamma}}\right]$$
\begin{equation}
\left\{(2p-p'-k)_{\nu}g_{\mu\rho}+(2p'-p-k)_{\rho}g_{\nu\mu}+(2k-p-p')_{\mu}g_{\rho\nu}\right\}
\end{equation}
After performing the dimensional regularization we can separate it
into two parts:planar and non-planar ones.
\begin{equation}
\Lambda^{(b)}_{\mu}(p,q,p')=\Lambda^{(b)}_{\mu}(p,q,p')_{planar}+\Lambda^{(b)}_{\mu}(p,q,p')_{nonplanar}
\end{equation}
(i)\hspace{0.5cm} For the planar part: \\
The complete expression of $ \Lambda^{(b)}_{\mu
(planar)}(p,q,p')$is given as:
$$\Lambda^{(b)}_{\mu(planar)}(p,q,p')=-\frac{2e^{2}}{(4\pi)^{\frac{d}{2}}}e^{\frac{i}{2}p\wedge
p'}\int_{0}^{1}dx\int_{0}^{1-x}dy\biggl\{(1-d)\gamma_{\mu}\left(\frac{\mu^{2}}{-M_{b}^{2}}\right)^{\frac{\epsilon}{2}}
\Gamma(\frac{\epsilon}{2})$$
\begin{equation}
-\frac{N_{2b}}{2M_{b}^{2}}\left(\frac{\mu^{2}}{-M_{b}^{2}}\right)^{\frac{\epsilon}{2}}
\Gamma(\frac{\epsilon}{2}+1)\biggr\}
\end{equation}
 In the limit
$d=4,\epsilon\rightarrow 0$, we get:
$$\Lambda^{(b)}_{\mu
(planar)}(p,q,p')=\left(\frac{3e^{2}}{8\pi^{2}}\right)\gamma_{\mu}e^{\frac{i}{2}p'\wedge
p}\left(\frac{1}{\epsilon}-\frac{1}{2}\gamma_{E}\right)$$
\begin{equation}
+\Lambda^{(b)\ast}_{\mu (planar)}(p,q,p')
\end{equation}
and $\Lambda^{(b)\ast}_{\mu (planar)}(p,q,p') $ is the finite part
of the vertex function for the planar diagram(b):
$$\Lambda^{(b)\ast}_{\mu
(planar)}(p,q,p')=-\frac{e^{2}}{16\pi^{2}}e^{\frac{i}{2}p\wedge
p'}\biggl\{\gamma_{\mu}\left(3\ln\frac{m^{2}}{\mu^{2}}+6\int_{0}^{1}dz
z\ln(1-z)^{2}\right)$$
\begin{equation}
-\int_{0}^{1}dx\int_{0}^{1-x}dy\frac{N_{2b}}{M^{2}_{b}}\biggr\}
\end{equation}
(ii)\hspace{0.5cm}For the non-planar part: \\ In the same way, the
contribution of the non-planar part is :
$$\Lambda^{(b)}_{\mu
(nonplanar)}(p,q,p')=\frac{2e^{2}\mu^{\epsilon}}{(4\pi)^{\frac{d}{2}}}e^{\frac{i}{2}p\wedge
p'}\int_{0}^{1}dx\int_{0}^{1-x}dy e^{i(x+y-1)(p\wedge p')}$$
$$\biggl\{\gamma_{\mu}\biggl[\frac{2(1-d)}{(M^{2}_{b})^{\frac{\epsilon}{2}}}
\left(\frac{Z_{b}}{2}\right)^{\frac{\epsilon}{2}}K_{\frac{\epsilon}{2}}(Z_{b})
-\frac{(\widetilde{q})^{2}}{2(M^{2}_{b})^{\frac{\epsilon}{2}-1}}
\left(\frac{Z_{b}}{2}\right)^{\frac{\epsilon}{2}-1}K_{\frac{\epsilon}{2}-1}(Z_{b})\biggr]$$
\begin{equation}
+\frac{(2-d)}{2}\frac{\widetilde{q}_{\mu}\widetilde{q\hspace{-0.20cm}/}}{(M^{2}_{b})^{\frac{\epsilon}{2}-1}}
\left(\frac{Z_{b}}{2}\right)^{\frac{\epsilon}{2}-1}K_{\frac{\epsilon}{2}-1}(Z_{b})
-\frac{{N_{2b}}}{(M^{2}_{b})^{\frac{\epsilon}{2}+1}}
\left(\frac{Z_{b}}{2}\right)^{\frac{\epsilon}{2}+1}K_{\frac{\epsilon}{2}+1}(Z_{b})\biggr\}
\end{equation}
 We also see that the divergent term exists only in the
planar part. Putting together (30) and (32) we receive the
contribution of the diagram (4b) :
$$\Lambda^{(b)}_{\mu}(p,q,p')=\Lambda^{(b)}_{\mu
(planar)}+\Lambda^{(b)}_{\mu (nonplanar)}(p,q,p')$$
\begin{equation}= \left(\frac{3e^{2}}{8\pi^{2}}\right)\gamma_{\mu}e^{\frac{i}{2}p'\wedge
p}\left(\frac{1}{\epsilon}-\frac{1}{2}\gamma_{E}\right)+\Lambda^{(b)\ast}_{\mu}(p,q,p')
\end{equation}
in which $ \Lambda^{(b)\ast}_{\mu}(p,q,p)$ is the finite part of
the contribution of diagram (4b) :
$$\Lambda^{(b)\ast}_{\mu}(p,q,p')=\Lambda^{(b)\ast}_{\mu
(planar)}(p,q,p')+\Lambda^{(b)}_{\mu(nonplanar)}(p,q,p')$$
$$=-\frac{e^{2}}{16\pi^{2}}e^{\frac{i}{2}p\wedge
p'}\biggl\{\gamma_{\mu}\left(3\ln\frac{m^{2}}{\mu^{2}}+6\int_{0}^{1}dz
z\ln(1-z)^{2}\right)-\int_{0}^{1}dx\int_{0}^{1-x}dy\frac{N_{2b}}{M^{2}_{b}}\biggr\}$$
$$+\frac{2e^{2}}{(4\pi)^{\frac{d}{2}}}e^{\frac{i}{2}p\wedge
p'}\int_{0}^{1}dx\int_{0}^{1-x}dy e^{i(x+y-1)(p\wedge p')}$$
$$\biggl\{\gamma_{\mu}\biggl[\frac{2(1-d)}{(M^{2}_{b})^{\frac{\epsilon}{2}}}
\left(\frac{Z_{b}}{2}\right)^{\frac{\epsilon}{2}}K_{\frac{\epsilon}{2}}(Z_{b})
-\frac{(\widetilde{q})^{2}}{2}\frac{1}{(M^{2}_{b})^{\frac{\epsilon}{2}-1}}
\left(\frac{Z_{b}}{2}\right)^{\frac{\epsilon}{2}-1}K_{\frac{\epsilon}{2}-1}(Z_{b})\biggr]$$
\begin{equation}
+\frac{(2-d)}{2}\frac{\widetilde{q}_{\mu}\widetilde{q\hspace{-0.20cm}/}}
{(M^{2}_{b})^{\frac{\epsilon}{2}-1}}\left(\frac{Z_{b}}{2}\right)^{\frac{\epsilon}{2}-1}K_{\frac{\epsilon}{2}-1}(Z_{b})
-\frac{{N_{2b}}}{(M^{2}_{b})^{\frac{\epsilon}{2}+1}}
\left(\frac{Z_{b}}{2}\right)^{\frac{\epsilon}{2}+1}K_{\frac{\epsilon}{2}+1}(Z_{b})\biggr\}
\end{equation}\\\\
 From (25) and(33) the contribution of $\psi\psi A$
vertex is given by:
\begin{equation}
\Lambda_{\mu}(p,q,p')=\left(\frac{3e^{2}}{8\pi^{2}}\right)\gamma_{\mu}e^{\frac{i}{2}p'\wedge
p}\left(\frac{1}{\epsilon}-\frac{1}{2}\gamma_{E}\right)+\Lambda_{\mu\ast}(p,q,p')
\end{equation}
where $\Lambda_{\mu\ast}(p,q,p)$ is the finite part of the total
vertex function and defined in (25) and (34) as:
\begin{equation}
\Lambda_{\mu}^{\ast}(p,q,p')=\Lambda^{(a)}_{\mu}(p,q,p')+\Lambda^{\ast(b)}_{\mu}(p,q,p')
\end{equation}
 As in ordinary QED ,at the one-loop level the total vertex for
the NCQED can be rewritten as:
$$\Gamma_{\mu}(p,q,p')=ie\left[(1+\Delta_{1})\gamma_{\mu}e^{\frac{i}{2}p\wedge p'}
+\Lambda_{\mu}^{div}+\Lambda_{\mu}^{\ast}\right]$$ where
$Z_{1}=1+\Delta_{1}$\\By choosing $$
\Delta_{1}=\Delta_{1}^{F}+\Delta_{1}^{\infty}$$ and imposing the
infinite renormalisation constant
$$\Delta_{1}^{\infty}=-\Lambda_{\mu}^{div}=-\frac{3e^{2}}{8\pi}\left(\frac{1}{\epsilon}-\frac{1}{2}\gamma_{E}\right)$$
we can remove the divergence part from the total vertex.\\
 The renormalisation constants $Z_{1}$ coming from the
one-loop vertex function is given by:
\begin{equation}
Z_{1}=1-\frac{3e^{2}}{8\pi^{2}\epsilon}
\end{equation}
At this stage we see that at the one-loop level the NCMQED is
completely renormalised.
\part*{{\large\bf 5. The $\beta$-function}}
As in the commutative QED, the
  $\beta$-function of the theory is given by:
  $$  \beta(g)=\mu\frac{\partial}{\partial\mu}g(\mu)$$
In terms of the bare coupling constant $g_{o}$ and the
  renormalization constants $Z_{i}$ ,(i=1,2,3), the renormalized
  coupling constant $g(\mu)$ is defined by:
  $$g_{o}=\mu^{\frac{\epsilon}{2}}g(\mu)Z_{1}Z_{2}^{-1}Z_{3}^{-\frac{1}{2}}$$
  where $\epsilon = 4-D$
Now bringing together(14),(23),(37)into the above relation,we get:
$$e_{o}=e\mu^{\frac{\epsilon}{2}}\left( 1-\frac{3e^{2}}{8\pi^{2}\epsilon}\right)
\left(1+\frac{e^{2}}{8\pi^{2}\epsilon}\right)\left[1-\frac{e^{2}}{8\pi^{2}\epsilon}
\left(\frac{20}{3}-\frac{8N_{f}}{3}\right)\right]$$
$$=e\mu^{\frac{\epsilon}{2}}\left[1+\frac{e^{2}}{16\pi^{2}\epsilon}\left(-\frac{22}{3}+\frac{4N_{f}}{3}\right)\right]$$
from which follows (in the limit $\epsilon\rightarrow 0$)
$$\beta(e)=-\frac{e^{3}}{16\pi^{2}}\left(\frac{22}{3}-\frac{4N_{f}}{3}\right)$$
Remarks: \\
-\hspace{1cm}A contribution $\frac{22}{3}$ is due to the structure
similar to non-abelian dynamics of NCM gauge fields.\\
-\hspace{1cm}From the evaluation, the UV divergence is suggested
to appear  only in the planar diagrams since the non-planar part
of the corresponding one-loop Feynman are assumed to be finite
with finite non-commutative parameter $\theta $\\
 -\hspace{1cm}For $ N_{f}<6$ the NCMQED with fundamental matters
 is asymptotically free.
\part*{{\large\bf 6.Vacuum polarisation of photon}}
Now, we determine the finite terms in the expression of
$i\Pi_{\mu\nu}^{(abc)}(p) $.Firstly,we evaluate
$i\Pi_{\mu\nu(non-planar)}^{(abc)}(p) $.In the  limit
$\epsilon\rightarrow 0,\widetilde{p}\rightarrow 0$ just keeping
the leader terms in the expansion of Bessel function we have:
 $$i\Pi_{\mu\nu(non-planar)}^{(abc)}(p)=\frac{ie^{2}}{(4\pi)^{2}}\int_{0}^{1}dz\biggl\{40\eta_{\mu\nu}
 \left[\frac{2}{|\widetilde{p}|^{2}}+M^{2}\ln\frac{Z}{2}-\frac{1}{2}M^{2}\right]$$
 $$+\left(-4\eta_{\mu\nu}|\widetilde{p}|^{2}
 +4\widetilde{p}_{\mu}\widetilde{p}\nu \right)\left(\frac{8}{|\widetilde{p}|^{4}}
 -\frac{2M^{2}}{|\widetilde{p}|^{2}}\right)+2\textit{M}_{\mu\nu}\ln\frac{Z}{2}\biggr\} $$
Taking integration over the Feynman parameter and putting together
with (20)we can rewrite the complete expression of
$i\Pi_{\mu\nu(p)}^{(abc)}(p) $ as:
$$i\Pi_{\mu\nu}^{(abc)}(p)=\frac{ie^{2}}{(4\pi)^{2}}\frac{10}{3}\left(\eta_{\mu\nu}p^{2}-p_{\mu}p_{\nu}\right)
\left(\frac{2}{\epsilon}+\ln
\pi|\widetilde{p}|^{2}\mu^{2}\right)$$
$$+\frac{ie^{2}}{(4\pi)^{2}}\biggl[\left(32\frac{\widetilde{p}_{\mu}\widetilde{p}_{\nu}}{|\widetilde{p}|^{4}}
-\frac{4}{3}p^{2}\frac{\widetilde{p}_{\mu}\widetilde{p}_{\nu}}{(\widetilde{p})^{2}}\right)\biggr]$$
We see that the vacuum polarisation of photon can be written in
the form:
$$i\Pi_{\mu\nu}^{(abc)}(p)=A\left(\eta_{\mu\nu}p^{2}-p_{\mu}p_{\nu}\right)+B\widetilde{p}_{\mu}\widetilde{p}_{\nu}$$
where A,B,and C are the functions of the scalars $ p^{2}$and $
|\widetilde{p}|^{2}$:
$$A=\frac{ie^{2}}{(4\pi)^{2}}\left( \frac{2}{\epsilon}+\ln
\pi|\widetilde{p}|^{2}\mu^{2}\right)$$
$$B=\frac{ie^{2}}{(4\pi)^{2}}\left(\frac{32}{(\widetilde{p})^{4}}-\frac{4}{3}\frac{p^{2}}{(\widetilde{p})^{2}} \right) $$
 That is the form compatible with the Ward identity.
 \section{THE FERMION-PHOTON VERTEX STRUCTURE IN NCMQED}
 Let's return to the electron-photon vertex contributions
from two diagrams (4a),(4b).\\Putting(26),(34)into(36)we get the
detail expression of the finite part for the proper vertex:
$$ \Lambda_{\mu}^{(\ast)}(p,q,p')=\frac{2e^{2}}{(4\pi)^{\frac{d}{2}}}e^{\frac{i}{2}p\wedge p'}\int_{0}^{1}dx\int_{0}^{1-x}dy
e^{-i(x+y)p\wedge p'}$$
$$\biggl\{\gamma_{\mu}\biggl[\frac{(2-d)^{2}}{2(M^{2}_{a})^{\frac{\epsilon}{2}}}
\left(\frac{Z_{a}}{2}\right)^{\frac{\epsilon}{2}}K_{\frac{\epsilon}{2}}(Z_{a})
-\frac{(2-d)}{4}\frac{(\widetilde{q})^{2}}{(M^{2}_{a})^{\frac{\epsilon}{2}-1}}
\left(\frac{Z_{a}}{2}\right)^{\frac{\epsilon}{2}-1}K_{\frac{\epsilon}{2}-1}(Z_{a})\biggr]$$
$$+\frac{(2-d)}{2}\frac{\widetilde{q}_{\mu}\widetilde{q\hspace{-0.20cm}/}}{(M^{2}_{a})^{\frac{\epsilon}{2}-1}}
\left(\frac{Z_{a}}{2}\right)^{\frac{\epsilon}{2}-1}K_{\frac{\epsilon}{2}-1}(Z_{a})
-\frac{N_{2a}}{(M^{2}_{a})^{\frac{\epsilon}{2}+1}}
\left(\frac{Z_{a}}{2}\right)^{\frac{\epsilon}{2}+1}K_{\frac{\epsilon}{2}+1}(Z_{a})\biggr\}$$
$$+\frac{2e^{2}}{(4\pi)^{\frac{d}{2}}}e^{\frac{i}{2}p\wedge
p'}\int_{0}^{1}dx\int_{0}^{1-x}dy e^{i(x+y-1)p\wedge p'}$$
$$\biggl\{\gamma_{\mu}\biggl[\frac{2(1-d)}{(M^{2}_{b})^{\frac{\epsilon}{2}}}
\left(\frac{Z_{b}}{2}\right)^{\frac{\epsilon}{2}}K_{\frac{\epsilon}{2}}(Z_{b})
-\frac{(\widetilde{q})^{2}}{2}\frac{1}{(M^{2}_{b})^{\frac{\epsilon}{2}-1}}
\left(\frac{Z_{b}}{2}\right)^{\frac{\epsilon}{2}-1}K_{\frac{\epsilon}{2}-1}(Z_{b})\biggr]$$
$$+\frac{(2-d)}{2}\frac{\widetilde{q}_{\mu}\widetilde{q\hspace{-0.20cm}/}}
{(M^{2}_{b})^{\frac{\epsilon}{2}-1}}\left(\frac{Z_{b}}{2}\right)^{\frac{\epsilon}{2}-1}K_{\frac{\epsilon}{2}-1}(Z_{b})
-\frac{{N_{2b}}}{(M^{2}_{b})^{\frac{\epsilon}{2}+1}}
\left(\frac{Z_{b}}{2}\right)^{\frac{\epsilon}{2}+1}K_{\frac{\epsilon}{2}+1}(Z_{b})\biggr\}$$
\begin{equation}
-\frac{e^{2}}{16\pi^{2}}e^{\frac{i}{2}p\wedge
p'}\biggl\{\gamma_{\mu}\left(3\ln\frac{m^{2}}{\mu^{2}}+6\int_{0}^{1}dz
z\ln(1-z)^{2}\right)-\int_{0}^{1}dx\int_{0}^{1-x}dy\frac{N_{2b}}{M^{2}_{b}}\biggr\}
\end{equation}
The structure of the electron-photon vertex in ordinary QED
indicates that the list of vectors and scalars appearing in the
vertex function was restricted to
$(\gamma_{\mu},q_{\mu},q^{2},m,e)$.In the case of NCQED,due to the
presence of $ \theta_{\mu\nu}$, we have two other scalars:
$\left((\widetilde{q})^{2},\widetilde{q\hspace{-0.20cm}}/\right)$,
and one other vector $(\widetilde{q}_{\mu} )$.Indeed,from the
above relation we can rewrite the vertex function in the form:
\begin{equation}
 \Lambda_{\mu}^{(\ast)}(p,q,p')=\left(\textit{G}\gamma_{\mu}+\textit{H}\widetilde{q}_{\mu}
 +\textit{L}\right)e^{\frac{i}{2}p\wedge p'}
 \end{equation}
 where$ \textit{G},\textit{H},\textit{L}$ are the functions of
 scalars $ (q^{2},(\widetilde{q})^{2},\widetilde{q\hspace{-0.20cm}}/,m,e)$.Theses functions  can be picked
 out from (38)in the on-shell condition:
$$\textit{G}(e,m,|\widetilde{q}|^{2})=\frac{2e^{2}}{(4\pi)^{\frac{d}{2}}}
\int_{0}^{1}dx\int_{0}^{1-x}dy\biggl\{e^{-i(x+y)p\wedge
p'}\biggl[\frac{(2-d)^{2}}{2(M^{2} _{a})^{\frac{\epsilon}{2}}}
\left(\frac{Z_{a}}{2}\right)^{\frac{\epsilon}{2}}K_{\frac{\epsilon}{2}}(Z_{a})$$
$$-\frac{(2-d)}{4}\frac{(\widetilde{q})^{2}}{(M^{2}_{a})^{\frac{\epsilon}{2}-1}}
\left(\frac{Z_{a}}{2}\right)^{\frac{\epsilon}{2}-1}K_{\frac{\epsilon}{2}-1}(Z_{a})\biggr]$$
$$ +e^{i(x+y-1)p\wedge p'}\biggl[\frac{2(1-d)}{(M^{2}_{b})^{\frac{\epsilon}{2}}}
\left(\frac{Z_{b}}{2}\right)^{\frac{\epsilon}{2}}K_{\frac{\epsilon}{2}}(Z_{b})
-\frac{(\widetilde{q})^{2}}{2}\frac{1}{(M^{2}_{b})^{\frac{\epsilon}{2}-1}}
\left(\frac{Z_{b}}{2}\right)^{\frac{\epsilon}{2}-1}K_{\frac{\epsilon}{2}-1}(Z_{b})\biggr]\biggr\}$$
\begin{equation}
-\frac{e^{2}}{(4\pi)^{\frac{d}{2}}}\biggl(3\ln\frac{m^{2}}{\mu^{2}}+6\int_{0}^{1}dz
z\ln(1-z)^{2}\biggr)
\end{equation}
$$\textit{H}( e,m,|\widetilde{q}|^{2},\widetilde{q\hspace{-0.2cm}/})=
\frac{(2-d)e^{2}\widetilde{q\hspace{-0.2cm}/}}{(4\pi)^{\frac{d}{2}}}\int_{0}^{1}dx\int_{0}^{1-x}dy
\biggl\{\frac{1}{(M^{2}_{a})^{\frac{\epsilon}{2}-1}}
\left(\frac{Z_{a}}{2}\right)^{\frac{\epsilon}{2}-1}K_{\frac{\epsilon}{2}-1}(Z_{a})e^{-i(x+y)p\wedge
p'}$$
\begin{equation}
+\frac{1}{(M^{2}_{b})^{\frac{\epsilon}{2}-1}}
\left(\frac{Z_{b}}{2}\right)^{\frac{\epsilon}{2}-1}K_{\frac{\epsilon}{2}-1}(Z_{b})e^{i(x+y-1)p\wedge
p'}\biggr\}
 \end{equation}\\\\
$$\textit{L}(e,m,|\widetilde{q}|^{2})=-\frac{2e^{2}}
 {(4\pi)^{\frac{d}{2}}}\int_{0}^{1}dx\int_{0}^{1-x}dy
\biggl\{\frac{{N_{2a}}}{(M^{2}_{a})^{\frac{\epsilon}{2}+1}}
\left(\frac{Z_{a}}{2}\right)^{\frac{\epsilon}{2}+1}K_{\frac{\epsilon}{2}+1}(Z_{a})e^{-i(x+y)p\wedge
p'}$$
\begin{equation}
+\frac{{N_{2b}}}{(M^{2}_{b})^{\frac{\epsilon}{2}+1}}\biggl[
\left(\frac{Z_{b}}{2}\right)^{\frac{\epsilon}{2}+1}K_{\frac{\epsilon}{2}+1}(Z_{b})e^{i(x+y-1)p\wedge
p'}-\frac{1}{2}\biggr] \biggr\}
\end{equation}
After sandwiching $\Gamma_{\mu}^{\ast}$ between $\overline{u}(p')$
and $ u(p)$ we obtain
\begin{equation}
\overline{u}(p')\Gamma_{\mu}^{\ast}u(p)=iee^{\frac{i}{2}p\wedge
p'}\overline{u}(p')\biggl[\textit{F}_{1}\gamma_{\mu}
+\frac{1}{2m}\textit{F}_{2}\left(i\sigma_{\mu\nu}q^{\nu}\right)
+\textit{H}\widetilde{q}_{\mu}\biggr]u(p)
\end{equation}
In the condition $q^{2}=0$ and in the limit $\epsilon\rightarrow
0$ the form factors are:
$$\textit{F}_{1}=1+\frac{2e^{2}}{(4\pi)^{2}}\int_{0}^{1}dx\int_{0}^{1-x}dy
\biggl\{\bigg[2K_{0}(Z_{a})e^{-i(x+y)p\wedge
p'}-6K_{0}(Z_{b})e^{i(x+y-1)p\wedge p'}\biggr]$$
$$+(\widetilde{q})^{2}\biggl[\frac{(M^{2}_{a})}{Z_{a}}K_{-1}(Z_{a})e^{-i(x+y)p\wedge
p'}-\frac{(M^{2}_{b})}{Z_{b}}K_{-1}(Z_{b})e^{i(x+y-1)p\wedge
p'}\biggr]\biggr\}$$
$$-\frac{e^{2}}{(4\pi)^{2}}\biggl(3\ln\frac{m^{2}}{\mu^{2}}+6\int_{0}^{1}dz
z\ln(1-z)^{2}\biggr)$$
$$+\frac{m^{2}e^{2}}{4\pi^{2}}\int_{0}^{1}dx\int_{0}^{1-x}dy
\biggl\{e^{-i(x+y)p\wedge p'} \frac{[(x+y+1)^{2}-3]}{M_{a}^{2}}
\left(\frac{Z_{a}}{2}\right)K_{1}(Z_{a})\biggr\}$$
\begin{equation}
+\frac{3m^{2}e^{2}}{4\pi^{2}}\int_{0}^{1}dx\int_{0}^{1-x}dy
\biggl[\left(\frac{Z_{b}}{2}\right)K_{1}(Z_{b})e^{i(x+y-1)p\wedge
p'}-\frac{1}{2}\biggr]
\end{equation}
and,
$$\textit{F}_{2}(q^{2}=0)=-\frac{8m^{2}e^{2}}{(4\pi)^{2}}\int_{0}^{1}dx\int_{0}^{1-x}dy$$
$$\biggl\{\frac{(x+y)(x+y-1)}{M_{a}^{2}}
\left(\frac{Z_{a}}{2}\right)K_{1}(Z_{a})e^{-i(x+y)p\wedge p'}$$
\begin{equation}+\frac{(x+y-1)(x+y-3)}{M_{b}^{2}}\biggl[ \left(\frac{Z_{b}}{2}\right)K_{1}(Z_{b})e^{i(x+y-1)p\wedge p'}
-\frac{1}{2}\biggr]\biggr\}
\end{equation}\\\\
We can see that (43) is the general form of the vertex which
satisfied the gauge invariance.\\As in the ordinary QED,the form
factor $\textit{F}_{1}$ is the correction of the fermion's charge.
At the tree level it's natural to impose:
$$\textit{F}_{1}(q^{2}=0) =1 $$
In this case, e is the electric charge of fermion in the limit
$q^{2}=0 $.From (44)this choice corresponds to
$(\epsilon\rightarrow 0)$:
$$\Delta_{1}^{F}=-\frac{2e^{2}}{(4\pi)^{2}}\int_{0}^{1}dx\int_{0}^{1-x}dy
\biggl\{\bigg[2K_{0}(Z_{a})e^{-i(x+y)p\wedge
p'}-6K_{0}(Z_{b})e^{i(x+y-1)p\wedge p'}\biggr]$$
$$+(\widetilde{q})^{2}\biggl[\frac{(M^{2}_{a})}{Z_{a}}K_{-1}(Z_{a})e^{-i(x+y)p\wedge
p'}-\frac{(M^{2}_{b})}{Z_{b}}K_{-1}(Z_{b})e^{i(x+y-1)p\wedge
p'}\biggr]\biggr\}$$
$$+\frac{e^{2}}{(4\pi)^{2}}\biggl(3\ln\frac{m^{2}}{\mu^{2}}+6\int_{0}^{1}dz
z\ln(1-z)^{2}\biggr)$$
$$-\frac{4m^{2}e^{2}}{(4\pi)^{2}}\int_{0}^{1}dx\int_{0}^{1-x}dy
\biggl\{e^{-i(x+y)p\wedge p'} \frac{[(x+y+1)^{2}-3]}{M_{a}^{2}}
\left(\frac{Z_{a}}{2}\right)K_{1}(Z_{a})\biggr\}$$
\begin{equation}
-\frac{12m^{2}e^{2}}{(4\pi)^{2}}\int_{0}^{1}dx\int_{0}^{1-x}dy
\frac{(x+y-1)^{2}}{M_{b}^{2}}
\biggl[\left(\frac{Z_{b}}{2}\right)K_{1}(Z_{b})e^{i(x+y-1)p\wedge
p'}-\frac{1}{2}\biggr]
\end{equation}
The form factor $F_{2}(q^{2}=0)$ is used to determine the
coefficient of the anomalous magnetic moment of fermion.
\section{THE WARD IDENTITY  }
Let us first consider the total vertex $ \Gamma_{\mu}(p,q,p')$.By
the counter terms $Z_{i}=1+\Delta_{i}$ we can rewrite the total
vertex in the form:
\begin{equation}
\Gamma_{\mu}(p,q,p')=ie\left[\gamma_{\mu}e^{\frac{i}{2}p\wedge
p'}(1+\Delta_{1})+\Lambda_{\mu}(p,q,p')\right]
\end{equation}
Recalling that:
$$\Lambda_{\mu}(p,q,p')=\Lambda_{\mu}^{a}(p,q,p')+\Lambda_{\mu}^{b}(p,q,p')$$
and,\\
$$\Lambda_{\mu}^{(a)}(p,q,p')
=-ie^{2}\mu^{\epsilon}e^{\frac{i}{2}p\wedge p'}\int
\frac{d^{d}k}{(2\pi)^{d}}e^{-ik\tilde{q}}\left[\gamma_{\nu}\frac{1}{D(p'-k)}\gamma_{\mu}
\frac{1}{D(p-k)}\gamma^{\nu}\frac{1}{k^{2}}\right]$$
$$\Lambda_{\mu}^{(b)}(p,q,p')
=+ie^{2}\mu^{\epsilon}e^{\frac{i}{2}p\wedge p'}\int
\frac{d^{d}k}{(2\pi)^{d}}\left(1-e^{ik\widetilde{q}}e^{-ip\wedge
p'}\right)$$
$$\left[\gamma^{\rho}\frac{1}{D(k)}
\gamma^{\nu}\frac{1}{(p'-k)^{2}}\frac{1}{(p-k)^{2}}\right]$$
\begin{equation}
\left\{(2p-p'-k)_{\nu}g_{\mu\rho}+(2p'-p-k)_{\rho}g_{\nu\mu}+(2k-p-p')_{\mu}g_{\rho\nu}\right\}
\end{equation}
where:
$$D(p)\equiv  p\hspace{-0.2cm}/-m $$
By the simple manipulations we can show that:
$$(p-p')^{\mu}\Lambda_{\mu}(p,q,p')=-ie^{2}\mu^{\epsilon}e^{\frac{i}{2}p\wedge
p'}\int\frac{d^{d}k}{(2\pi)^{d}}
\left(e^{ik\widetilde{q}}e^{-ip\wedge
p'}-1\right)\frac{\textit{N}_{f}}{\textit{D}}$$
\begin{equation}
-ie^{2}\mu^{\epsilon}e^{\frac{i}{2}p\wedge
p'}\int\frac{d^{d}k}{(2\pi)^{d}} \left[
\frac{1}{k^{2}}\gamma^{\rho}\frac{1}{D(p'-k)}\gamma_{\rho}
-\frac{1}{k^{2}}\gamma^{\rho}\frac{1}{D(p-k)}\gamma_{\rho} \right]
\end{equation}
where:
\begin{equation}
\textit{N}_{f}=2(p\hspace{-0.2cm}/-p\hspace{-0.2cm}/')+2(p-p')kp\hspace{-0.2cm}/'
-2k^{2}(p\hspace{-0.2cm}/-p\hspace{-0.2cm}/')-2mk(p-p')
\end{equation}Recalling that the expression of the fermion self energy is:
\begin{equation}
\sum_{(p)}=-\Delta_{2}p\hspace{-0.2cm}/-\Delta_{0}m+\widetilde{\sum}_{(p)}
\end{equation}
 in which $\Delta_{2}$ and $\Delta_{0}$ are the
counter terms for the fermion loop while $\widetilde{\sum}_{(p)}$
is the correction of the fermion's propagator whose expression is:
\begin{equation}
\widetilde{\sum}_{(p)}=-ie^{2}\mu^{\epsilon}e^{\frac{i}{2}p\wedge
p'}\int\frac{d^{d}k}{(2\pi)^{d}}\gamma^{\rho}\frac{1}{D(p-k)}\gamma_{\rho}\frac{1}{k^{2}}
\end{equation}
From (49) and (52) we obtain the relation between the
$\lambda_{\mu}'s$ and the $\widetilde{\sum}_{(p)}'s $ as:
\begin{equation}
(p-p')^{\mu}\Lambda_{\mu}(p,q,p')= e^{\frac{i}{2}p\wedge
p'}\left(\widetilde{\sum}_{(p')}-\widetilde{\sum}_{(p)}+\Omega\right)
\end{equation}
 The $\Omega $'s term in the above equation is defined
as:
\begin{equation}
\Omega\equiv -ie^{2}\mu^{\epsilon}\int
\frac{d^{d}k}{(2\pi)^{d}}\left(e^{ik\widetilde{q}}e^{-ip\wedge
p'}-1\right)\frac{\textit{N}_{f}}{\textit{D}}
\end{equation}
where:
\begin{equation}
N_{f} =2pk(p\hspace{-0.2 cm}/-p\hspace{-0.2 cm}/')+2p\hspace{-0.2
cm}/'(p-p')k-2k^{2}(p\hspace{-0.2 cm}/-p\hspace{-0.2
cm}/')-2mk(p-p')
\end{equation}
By the standard Feynman parameterisation , it is easy to show that
$\Omega$ can be decomposed into two parts: finite and infinite .
\begin{equation}
\Omega=(p\hspace{-0.2
cm}/-p\hspace{-0.2cm}/')\left(\Omega^{\infty}+\Omega^{F}\right)
\end{equation}
in which we have:
\begin{equation}
\Omega^{\infty}(q^{2}=0)=\frac{e^{2}}{4\pi^{2}}\left(\frac{1}{\epsilon}-\frac{1}{2}\gamma_{E}\right)
\end{equation}
and;
$$\Omega^{F}(q^{2}=0)=-\frac{e^{2}}{8\pi^{2}}\ln\frac{m^{2}}{\mu^{2}}-\frac{e^{2}}{4\pi^{2}}\int_{0}^{1}dzz\ln(1-z)^{2}
+\frac{e^{2}}{8\pi^{2}}\int_{0}^{1}\frac{z^{2}}{(1-z)}$$
$$-\frac{e^{2}}{4\pi^{2}}\int_{0}^{1}dx\int_{0}^{1-x}dy
e^{i(x+y-1)p\wedge p'}$$
\begin{equation}\biggl\{2K_{0}(Z_{b})+\frac{(\widetilde{q})^{2}}{2}\frac{M_{b}^{2}}{Z_{b}}K_{-1}(Z_{b})
+\frac{m^{2}(x+y)(1-x-y)}{M_{b}^{2}}\left(\frac{Z_{b}}{2}\right)K_{1}(Z_{b})\biggr\}
\end{equation}
Now,returning to (47),from (51)and (53) we obtain:
\begin{equation}
(p-p')^{\mu}\Gamma_{\mu}(p,q,p')=iee^{\frac{i}{2}p\wedge
p'}\biggl\{(p\hspace{-0.2cm}/-p\hspace{-0.2cm}/')\left[\Delta_{1}-\Delta_{2}+\Omega^{\infty}+\Omega^{F}\right]
+S^{-1}(p)-S^{-1}(p')\biggr\}
\end{equation}
For the divergence parts ,with the result (13) and (37) in
section.3:
$$\Delta_{2}^{\infty}=-\frac{e^{2}}{16\pi^{2}}\left(\frac{2}{\epsilon}-\gamma_{E}\right) $$
$$\Delta_{1}^{\infty}=-\frac{3e^{2}}{8\pi}\left(\frac{1}{\epsilon}-\frac{1}{2}\gamma_{E}\right) $$
fortunately,we see that :
$$\Delta_{1}^{\infty}-\Delta_{2}^{\infty}+\Omega^{\infty}=0$$
and we obtain:
\begin{equation}
(p-p')^{\mu}\Gamma_{\mu}(p,q,p')=iee^{\frac{i}{2}p\wedge
p'}(p\hspace{-0.2cm}/-p\hspace{-0.2cm}/')\left[\Delta_{1}^{F}-\Delta_{2}^{F}+\Omega^{F}\right]+iee^{\frac{i}{2}p\wedge
p'}\left[S^{-1}(p)-S^{-1}(p')\right]
\end{equation}
For the finite part, from (13),(46)and (58)we see that:
$$\Delta_{1}^{F}-\Delta_{2}^{F}+\Omega^{F}\neq 0$$
That means we can redefine the charge of fermion e .
\section{THE ANOMALOUS MAGNETIC MOMENT}
As discussed in [5] ,we see that the coefficient $\textit{H}$ in
the expression of vertex function (43) will give the new
contribution to the magnetic moment:
\begin{equation}
<\overrightarrow{\mu}>=\frac{\textit{H}}{i}\overrightarrow{\theta},
\hspace{2cm}\theta_{i}\equiv\epsilon_{ijk}\theta_{jk}
\end{equation}
As we see this correction of magnetic moment does not depend on
spin.So far we can write the magnetic moment of fermion in NCQED
as:
\begin{equation}
<\overrightarrow{\mu}>=g\left(\frac{e}{2m}\right)\overrightarrow{s}+\frac{\textit{H}}{i}\overrightarrow{\theta}
\end{equation}
The form factor $\textit{F}_{2}(q^{2}=0) $ determines the
coefficient of the anomalous magnetic moment :
\begin{equation}
g=2[1+\textit{F}_{2}(q^{2}=0)]
\end{equation}
Now from (45) we can identify $\textit{F}_{2}(q^{2}=0)$ as:
$$\textit{F}_{2}(q^{2}=0)=-\frac{8m^{2}e^{2}}{(4\pi)^{2}}\int_{0}^{1}dx\int_{0}^{1-x}dy$$
$$\biggl\{\frac{(x+y)(x+y-1)}{M_{a}^{2}}
\left(\frac{Z_{a}}{2}\right)K_{1}(Z_{a})e^{-i(x+y)p\wedge p'}$$
\begin{equation}+\frac{(x+y-1)(x+y-3)}{M_{b}^{2}}\biggl[ \left(\frac{Z_{b}}{2}\right)K_{1}(Z_{b})e^{i(x+y-1)p\wedge p'}
-\frac{1}{2}\biggr]\biggr\}
\end{equation}\\\\
Now, we try to evaluate the magnetic moment of fermion in the rest
framework such that $p\wedge p'=0 $ in which for $ \theta^{0i}= 0$
we get:
$$\textit{F}_{2}(q^{2}=0)=-\frac{8m^{2}e^{2}}{(4\pi)^{2}}\int_{0}^{1}dx\int_{0}^{1-x}dy$$
\begin{equation}
\biggl\{\frac{(x+y)(1-x-y)}{M_{a}^{2}}
\left(\frac{Z_{a}}{2}\right)K_{1}(Z_{a})
+\frac{(x+y-1)(x+y-3)}{M_{b}^{2}}\biggl[
\left(\frac{Z_{b}}{2}\right)K_{1}(Z_{b})
-\frac{1}{2}\biggr]\biggr\}
\end{equation}
Now in the limit $ |\widetilde{q}|\approx 0$  by keeping the
leader terms in the expansion of the Bessel function , we obtain:
$$\textit{F}_{2}(q^{2}=0)=-\frac{8m^{2}e^{2}}{(4\pi)^{2}}\int_{0}^{1}dx\int_{0}^{1-x}dy$$
$$\biggl\{(x+y)(1-x-y)
\biggl[\frac{1}{2M^{2}_{a}}+\frac{|\widetilde{q}|^{2}}{8}\ln\frac{m^{2}|\widetilde{q}|^{2}}{4}
+\frac{|\widetilde{q}|^{2}}{8}\ln\left[(x+y)^{2}+\frac{m_{\gamma}^{2}}{m^{2}}(1-x-y)\right]$$
$$+\frac{|\widetilde{q}|^{2}}{4}(-\frac{1}{2}+\gamma_{E})\biggr]$$
$$+(x+y-1)(x+y-3)\biggl[\frac{|\widetilde{q}|^{2}}{8}\ln\frac{m^{2}|\widetilde{q}|^{2}}{4}
+\frac{|\widetilde{q}|^{2}}{8}\ln\left[(x+y-1)^{2}+\frac{m^{2}_{\gamma}}{m^{2}}(x+y)\right]$$
\begin{equation}
+\frac{|\widetilde{q}|^{2}}{4}\left(-\frac{1}{2}+\gamma_{E}\right)\biggr]\biggr\}
\end{equation}
After the integration over the Feynman parameters we get:
$$\textit{F}_{2}(q^{2}=0)=\frac{\alpha}{2\pi}
-\frac{m^{2}e^{2}|\widetilde{q}|^{2} }{48\pi^{2}}$$
\begin{equation}\biggl\{\ln\frac{m^{2}|\widetilde{q}|^{2}}{4}-1+2\gamma_{E}
+6\int_{0}^{1}z(1-z)\ln[z^{2}+\frac{m_{\gamma}^{2}}{m^{2}}(1-z)]\biggr\}
\end{equation}
Due to the massless photon  there is an IR-divergent term in the
expression of the form factor $F_{2}(q^{2})=0$.Ignoring the
IR-divergent terms we receive the correction of the coefficient g
as:
\begin{equation}
 \delta g=-\frac{m^{2}e^{2}|\widetilde{q}|^{2} }{48\pi^{2}}
\biggl\{\ln\frac{m^{2}|\widetilde{q}|^{2}}{4}-1+2\gamma_{E}\biggr\}
\end{equation}
Now, we evaluate the contribution of the coefficient $\textit{H}$
in the rest frame.From (41) we have:
\begin{equation}\textit{H}(e,m,|\widetilde{q}|^{2},\widetilde{q\hspace{-0.2cm}/})=
-\frac{4e^{2}\widetilde{q\hspace{-0.2cm}/}}{(4\pi)^{2}}\int_{0}^{1}dx\int_{0}^{1-x}dy
\biggl\{\frac{(M^{2}_{a})}{Z_{a}}K_{-1}(Z_{a})
+\frac{(M^{2}_{b})}{Z_{b}}K_{-1}(Z_{b})\biggr\}
 \end{equation}\\\\
In the low momentum limit,keeping the leader terms in the
expansion of Bessel's function we get:
\begin{equation}
\textit{H}=-\frac{e^{2}(\widetilde{q\hspace{-0.2cm}/})}{4\pi^{2}|\widetilde{q}|^{2}}
\end{equation}\\\\
So, the full expression of the fermion's magnetic moment  in the
rest frame is:
\begin{equation}
<\overrightarrow{\mu}>=\frac{e}{m}\biggl\{1+\frac{\alpha}{2\pi}
-\frac{m^{2}e^{2}|\widetilde{q}|^{2} }{48\pi^{2}}
\biggl[\ln\frac{m^{2}|\widetilde{q}|^{2}}{4}-1+2\gamma_{E}\biggr]\biggr\}\overrightarrow{s}
+\frac{ie^{2}(\widetilde{q\hspace{-0.2cm}/})}{4\pi^{2}|\widetilde{q}|^{2}}\overrightarrow{\theta}
\end{equation}

\section{\Large\bf CONCLUSIONS}
In this paper we have studied the renormalisation of NCMQED ,the
vacuum polarisation of photon, the$ \beta$-function,the
contribution of the vertex function at one-loop level in NCQED
.Based on the dimensional regularization method which have been
generalized to NC-theories and by assuming Bessel functions are
finite at the $\theta $ finite the calculations show that the
theory is renormalised with the counter terms.The structure of
vacuum polarisation of photon satisfies Ward identity as well as
the structure of vertex .It is shown that besides the normal form
factors there is a new contribution to the magnetic moment comes
from the parameter $\theta$. The Ward identity is satisfied at the
one-loop level of vertex and in the condition $\theta^{0\nu}=0 $.
\part*{{\Large\bf ACKNOWLEDGMENT}}
We would like to thank Professor J.P Derendinger for
encouragement,kind attention and many helpful discussions.
\pagebreak \part*{{\Large\bf Appendix}} We work in d-dimensional '
"Minkowski" space with one timelike and (d-1) spacelike
dimensions.We are interested in the generic
integrals : \\
$$(2.1)\hspace{1cm}I_{1}=\int\frac{d^{D}k}{(2\pi)^{D}}\frac{e^{ik\tilde{q}}}{(k^{2}-\Delta)^{3}}$$
Using Wick's rotation:$k^{0}=ik^{n}$, we have:
$$ k^{2}=-k^{2}_{E}$$
and,\\
$$ k\widetilde{q}=k^{\mu}\theta^{\mu\nu}q_{\nu}=k^{0}\theta^{0\nu}q_{\nu}+k^{i}\theta^{i\varepsilon}q_{\varepsilon}$$
In the case of the spatial non-commutativity $\theta^{0\nu}=0$ \\
$$ \Rightarrow k\widetilde{q}=k^{i}\theta^{ij}q_{j}$$ So,the exponent
$ e^{ikq}$ doesn't change the sign and we get
$$I_{1}=(i)(-1)^{3}\int\frac{d^{D}k}{(2\pi)^{D}}\frac{e^{ik\tilde{q}}}{(k^{2}+\Delta)^{3}}=(i)(-1)^{3}I_{1E}$$
In order to evaluate the generic form of this integral in
Euclidean space we use the Feynman parameterisation:
$$ \frac{1}{k^{2}+\Delta}=\int_{0}^{\infty} d\alpha
e^{-\alpha(k^{2}+\Delta)}$$
$$\Rightarrow I_{1E}=\int_{0}^{\infty}d\alpha_{1}d\alpha_{2}d\alpha_{3}\int\frac{d^{D}k}{(2\pi)^{D}}
e^{-(\alpha_{1}+\alpha_{2}+\alpha_{3})k^{2}+ik\tilde{q}-\Delta(\alpha_{1}+\alpha_{2}+\alpha_{3})}$$
Taking the integration over k we get
$$I_{1E}=\frac{1}{(4\pi)^{\frac{D}{2}}}\int_{0}^{\infty}d\alpha_{1}d\alpha_{2}d\alpha_{3}
\frac{1}{(\alpha_{1}+\alpha_{2}+\alpha_{3})^{\frac{D}{2}}}
exp[-\frac{(\tilde{q})^{2}}{4(\alpha_{1}+\alpha_{2}+\alpha_{3})}-\Delta(\alpha_{1}+\alpha_{2}+\alpha_{3})]$$
Inserting
$$ 1=\int_{0}^{\infty} d\rho \cdot
\delta(\rho-\sum_{i=1}^{3}\alpha_{i})$$ and rescaling
$\alpha_{i}\rightarrow\rho\alpha_{i}$ we get
$$I_{1E}=\frac{1}{(4\pi)^{\frac{D}{2}}}\frac{1}{\Gamma}(3)\int_{0}^{\infty}\frac{d\rho}{\rho^{\frac{D}{2}-2}}
exp[-\frac{(\tilde{q})^{2}}{4\rho}-\Delta\rho] $$ In terms of
Bessel function we have:
$$I_{1E}=\frac{1}{(4\pi)^{\frac{D}{2}}}\frac{1}{\Gamma(3)}\frac{1}{\Delta^{3-\frac{D}{2}}}
2\left(\frac{Z}{2}\right)^{3-\frac{D}{2}}K_{3-\frac{D}{2}}(Z)$$
where $ Z=|\tilde{q}|(\Delta)^{\frac{1}{2}}$ and,
$$ I_{1}=(i)(-1)^{3}\frac{1}{(4\pi)^{\frac{D}{2}}}\frac{1}{\Gamma(3)}\frac{1}{\Delta^{3-\frac{D}{2}}}
2\left(\frac{Z}{2}\right)^{3-\frac{D}{2}}K_{3-\frac{D}{2}}(Z)$$
$$(2.2)\hspace{1cm} I_{\mu\nu}=\int\frac{d^{D}k}{(2\pi)^{D}}\frac{k_{\mu}k_{\nu}e^{ik\tilde{q}}}{(k^{2}-\Delta)^{3}}$$
After the Wick rotation the integral has the form:
$$I_{\mu\nu}=(i)(-1)^{3}I_{\mu\nu(E)}=(i)(-1)^{3}\int\frac{d^{D}k}{(2\pi)^{D}}
\frac{k_{\mu}k_{\nu}e^{ik\tilde{q}}}{(k^{2}+\Delta)^{3}} $$ We
evaluate $ I_{\mu\nu(E)}$ from the generating function:
$$Z_{E}=\int\frac{d^{D}k}{(2\pi)^{D}}\frac{e^{ik(\tilde{q}-z)}}{(k^{2}+\Delta)^{3}}$$
we see that:
$$I_{\mu\nu(E)}=\frac{1}{i^{2}}\frac{\partial^{2}}{\partial z_{\mu}\partial
z_{\nu}}Z|_{z\rightarrow0}$$ As the same way, after the
parameterisation, we have:
 $$ Z_{E}=\int_{0}^{\infty}d\alpha_{1}d\alpha_{2}d\alpha_{3}\int\frac{d^{D}k}{(2\pi)^{D}}
e^{-(\alpha_{1}+\alpha_{2}+\alpha_{3})k^{2}+ik(\tilde{q}-z)-\Delta(\alpha_{1}+\alpha_{2}+\alpha_{3})}$$
with the aid of the Gaussian integral we can take integration over
momentum k:
$$ Z_{E}=\frac{1}{(4\pi)^{\frac{D}{2}}}\int_{0}^{\infty}d\alpha_{1}d\alpha_{2}d\alpha_{3}
\frac{1}{(\alpha_{1}+\alpha_{2}+\alpha_{3})^{\frac{D}{2}}}
exp[-\frac{(\tilde{q}-z)^{2}}{4(\alpha_{1}+\alpha_{2}+\alpha_{3})}-\Delta(\alpha_{1}+\alpha_{2}+\alpha_{3})]$$
Now taking derivative with respect to z and then taking the
limit$z\rightarrow 0 $ we get:
$$I_{\mu\nu(E)}=\frac{1}{(4\pi)^{\frac{D}{2}}}\int_{0}^{\infty}d\alpha_{1}d\alpha_{2}d\alpha_{3}
\left\{\frac{\delta_{\mu\nu}}{2(\alpha_{1}+\alpha_{2}+\alpha_{3})^{\frac{D}{2}+1}}
-\frac{\widetilde{q}_{\mu}\widetilde{q}_{\nu}}{4(\alpha_{1}+\alpha_{2}+\alpha_{3})^{\frac{D}{2}+2}}\right\}$$
$$exp[-\frac{(\tilde{q})^{2}}{4(\alpha_{1}+\alpha_{2}+\alpha_{3})}-\Delta(\alpha_{1}+\alpha_{2}+\alpha_{3})]$$
Inserting
$$ 1=\int_{0}^{\infty} d\rho \cdot
\delta(\rho-\sum_{i=1}^{3}\alpha_{i})$$ and rescaling
$\alpha_{i}\rightarrow\rho\alpha_{i}$ we get
$$I_{\mu\nu(E)}=\frac{1}{(4\pi)^{\frac{D}{2}}}\frac{1}{\Gamma(3)}\int_{0}^{\infty} d\rho
\left\{\frac{\delta_{\mu\nu}}{2(\rho)^{\frac{D}{2}-1}}
-\frac{\widetilde{q}_{\mu}\widetilde{q}_{\nu}}{4(\rho)^{\frac{D}{2}}}\right\}$$
$$exp[-\frac{(\tilde{q})^{2}}{4(\rho)}-\Delta(\rho)]$$
In terms of the Bessel's functions,the relation above can be
rewritten in the form:
$$I_{\mu\nu(E)}=A\delta_{\mu\nu}-B\tilde{q}_{\mu}\tilde{q}_{\nu}$$
where
$$A=\frac{1}{(4\pi)^{\frac{D}{2}}}\frac{1}{\Gamma(3)}
\frac{1}{\Delta^{2-\frac{D}{2}}}\left(\frac{Z}{2}\right)^{2-\frac{D}{2}}K_{2-\frac{D}{2}}(Z)$$
$$B=\frac{1}{(4\pi)^{\frac{D}{2}}}\frac{1}{\Gamma(3)}\frac{1}{2\Delta^{1-\frac{D}{2}}}
\left(\frac{Z}{2}\right)^{1-\frac{D}{2}}K_{1-\frac{D}{2}}(Z)$$
returning to the Minkowski space, the result is:
$$ I_{\mu\nu)}=(-i)(-1)^{3}\left[Ag_{\mu\nu}+B\tilde{q}_{\mu}\tilde{q}_{\nu}\right]$$
In general,the integral:
$$ (2.2)\hspace{1cm} I_{\mu\nu}=\int\frac{d^{D}k}{(2\pi)^{D}}
\frac{k_{\mu}k_{\nu}e^{ik\tilde{q}}}{(k^{2}-\Delta)^{\alpha}}
=(-i)(-1)^{\alpha}\left[Ag_{\mu\nu}+B\tilde{q}_{\mu}\tilde{q}_{\nu}\right]$$
where:
$$A=\frac{1}{(4\pi)^{\frac{D}{2}}}\frac{1}{\Gamma(\alpha)}
\frac{1}{\Delta^{\alpha-1-\frac{D}{2}}}\left(\frac{Z}{2}\right)^{\alpha-1-\frac{D}{2}}K_{\alpha-1-\frac{D}{2}}(Z)$$
$$B=\frac{1}{(4\pi)^{\frac{D}{2}}}\frac{1}{\Gamma(\alpha)}\frac{1}{2}\frac{1}{\Delta^{\alpha-2-\frac{D}{2}}}
\left(\frac{Z}{2}\right)^{\alpha-2-\frac{D}{2}}K_{\alpha-2-\frac{D}{2}}(Z)$$
$$(2.3)\hspace{1cm}I_{2}=\int\frac{d^{D}k}{(2\pi)^{D}}\frac{k^{2}e^{ik\tilde{q}}}{(k^{2}-\Delta)^{\alpha}}$$
Contracting the result is : $$I_{2}=(-i)(-1)^{\alpha}[A\cdot
D+B(\tilde{q})^{2}]$$
\newpage
\section*{References}
\begin{enumerate}
\item \label{}N.Seiberg and E.Witten,{\it String Theory and
Noncommutative Geometry}  J.High Energy
Phys.09(1999)032,hep-th/9908142 \item
\label{}S.Minwalla,M.Van.Raamsdonk,N.Seiberg,{\it Noncommutative
Perturbative Dynamics},hep-th /9912072 \item
\label{}C.P.Martin,D.Sanchez.Ruiz,Phys.Rev.Letter.83(1999)476.
\item \label{}M.Hayakawa,{\it Perturbative analysis on infrared
and ultraviolet aspects of noncommutative QED on
$R^{4}$},Phys.Letter.B478(2000)394,hep-th/9912167 \item
\label{}I.F.Riad and M.M.Sheikh.Jabbari,{\it Noncommutative QED
and Anomalous Dipole Moment},J.High Energy
Phys.08(2000)045,hep-th/0008132 \item
\label{}F.Ardalan,N.Sadooghi,{\it Axial Anomaly in Non-commutative
QED on $R^{4}$},hep-th/0002143 \item
\label{}F.T.Brandt,Ashok.Das,J.Frenkel,{\it General Structure of
the Photon Self Energy in NCQED},hep-th/0112127  \item
\label{}Xiao-Jun Wang,Mu-LiYan, {\it Noncommutative QED and Muon
Anomaluos Magnetic Moment},hep-th/0109095. \item
\label{}A.Matusis,L.Susskind,N.Toumbas,{\it The IR/UV Connection
in the Noncommutative Gauge Theories},hep-th/0002075.D.Bigatti and
L.Susskind,{\it Magnetic fields,branes and noncommutative
geometry},hep-th/9908056.  \item \label{}Tadahito NAKAJIMA,{\it
Conformal Anomalies in Noncommutative Gauge
Theories},hep-th/0108158. \item \label{}Neda Sadooghi and Mojtaba
Mohammadi,{\it On the beta-function and conformal Anomaly of
Noncommutative QED with Adjoint Matter Fields},hep-th/0206137
\item \label{}X.Calmet,B.Jurco,P.Schupp,J.Wess,and
M.Wohlgenannt,{\it The standard model on Noncommutative
space-time},Eur.Phys.J.C23 (2002)363 \item \label{}Schwarz,A.{\it
Noncommutative instanton:A new approach},hep-th/0102182 \item
\label{}Michael R Douglas and Nikita A.Nekrasov,{\it
Noncommutative Field theory},hep-th/0106048\item \label{}Katsusada
Morita,{\it Lorentz-Invariant Noncommutative QED},hep-th/0209234
\item \label{}M.Chaichian,A.Demichev,P.Presnajder,A.Tureanu,{\it
Space-Time Noncommutativity,Discreteness of Time and
Unitarity},hep-th/0007156. \item \label{}.M.M.Sheikh-Jabbari,{\it
Discrete Symmetries (C,P,T) in Noncommutative Field
Theories},Phys.Rev.Lett.84(2000)5265,hep-th/000167. \item \label{}
Ki Boum Eom,Sung-Shig Kang,Bum-Hoon Lee,Chanyong Park,{\it
Radiative Corrections in Noncommutative QED},hep-th/0205093. \item
\label{}C.P Martin and D.Sanchez-Ruiz,{\it The one loop UV
Divergent Structure of U(1) Yang-Mills Theory on Noncommutative
$R^{4}$},Phys.Rev.Lett.83 (1999) 476. \item \label{}J.Gomis and
T.Mehen,{\it Space-Time Noncommutative Field Theories and
Unitary},hep-th/0005129.\item \label{}T.Mariz,C.A.,de S.Pires and
R.F.Ribeiro,{\it Ward identity in Noncommutative
QED},hep-th/0211416. \item \label{}J-P.Derendinger,{\it Theorie
quantique des champs }\item \label{}C.Itzykson and J-B.Zuber,{\it
Quantum Field Theory},McGraw-Hill,1985  \item \label{}M.E.Peskin
and D.v.Schroeder,{\it An Introduction to Quantum Field
Theory},Addison-Wesley Publishing Company,1995.
\end{enumerate}
\normalsize
\end{document}